# Learning-based Application-Agnostic 3D NoC Design for Heterogeneous Manycore Systems

Biresh Kumar Joardar, *Student Member*, *IEEE*, Ryan Gary Kim 🆔, *Member*, *IEEE*, Janardhan Rao Doppa 🆔, *Member*, *IEEE*, Partha Pratim Pande 🆔, *Senior Member*, *IEEE*, Diana Marculescu, *Fellow*, *IEEE*, and Radu Marculescu 🆔, *Fellow*, *IEEE*

**Abstract**— The rising use of deep learning and other big-data algorithms has led to an increasing demand for hardware platforms that are computationally powerful, yet energy-efficient. Due to the amount of data parallelism in these algorithms, high-performance three-dimensional (3D) manycore platforms that incorporate both CPUs and GPUs present a promising direction. However, as systems use heterogeneity (*e.g.*, a combination of CPUs, GPUs, and accelerators) to improve performance and efficiency, it becomes more pertinent to address the distinct and likely conflicting communication requirements (*e.g.*, CPU memory access latency or GPU network throughput) that arise from such heterogeneity. Unfortunately, it is difficult to quickly explore the hardware design space and choose appropriate tradeoffs between these heterogeneous requirements. To address these challenges, we propose the design of a 3D Network-on-Chip (NoC) for heterogeneous manycore platforms that considers the appropriate design objectives for a 3D heterogeneous system and explores various tradeoffs using an efficient machine learning (ML-)based multi-objective optimization (MOO) technique. The proposed design space exploration considers the various requirements of its heterogeneous components and generates a set of 3D NoC architectures that efficiently trades off these design objectives. Our findings show that by *jointly* considering these requirements (latency, throughput, temperature, and energy), we can achieve 9.6% better Energy-Delay Product on average at nearly iso-temperature conditions when compared to a thermally-optimized design for 3D heterogeneous NoCs. More importantly, our results suggest that our 3D NoCs optimized for a few applications can be generalized for unknown applications as well. Our results show that these generalized 3D NoCs only incur a 1.8% (36-tile system) and 1.1% (64-tile system) average performance loss compared to application-specific NoCs.

————————————  ◆  ————————————

## 1 INTRODUCTION

Neural Networks, graph analytics, and other big-data applications have become vastly important for many domains. This has led to a search for proper computing systems that can efficiently utilize the tremendous amount of data parallelism that is associated with these applications. Recently, platforms using both CPUs and GPUs have significantly improved the execution time for such applications [1]. However, existing discrete GPU systems use off-chip interconnects (*e.g.*, PCIe) to communicate with the CPUs. These interconnects give rise to high data transfer latency and become performance bottlenecks for applications that involve high volumes of data transfers between the CPUs and GPUs.

A heterogeneous manycore system that integrates many CPUs and GPUs on a single chip can solve this problem and avoid such expensive off-chip data transfers [2], [3]. In addition, these single-chip systems require a scalable interconnection backbone (Networks-on-Chip (NoCs)) to facilitate more efficient communication.

To further reduce data transfer costs, three-dimensional (3D) integrated circuits (ICs) have been investigated as a possible solution and have made significant strides towards improving communication efficiency [4], [5]. By connecting planar dies stacked on top of each other with through-silicon vias (TSVs), the communication latency, throughput, and energy consumption can be further improved [6].

3D ICs together with NoCs, enable the design of highly integrated heterogeneous (*e.g.*, CPUs, GPUs, accelerators) manycore platforms for big-data applications. However, the design of 3D NoC based manycore systems pose unique challenges. Due to the heterogeneity of the cores integrated on a single chip, the communication requirements for each core can vary significantly. For example, in a CPU-GPU based heterogeneous system, CPUs require low memory latency while GPUs need high-throughput data transfers [7]. In addition to the individual core requirements, 3D ICs allow dense circuit integration but have much higher power density than their 2D counterparts. Therefore, the design process must consider reducing temperature hotspots as an additional objective. Overall, the design of a 3D heterogeneous manycore architecture needs to consider each of these objectives and satisfy all of them simultaneously [8]. Hence, 3D heterogeneous manycore design can be formulated as a multi-objective

————————————————

- *Biresh Kumar Joardar, Janardhan Rao Doppa, and Partha Pratim Pande are with Washington State University, Pullman, WA, 99164. Email: {biresh.joardar, jana.doppa, pande}@wsu.edu*
- *Ryan Gary Kim is with Colorado State University, Fort Collins, CO, 80523, Email: Ryan.G.Kim@colostate.edu*
- *Diana Marculescu and Radu Marculescu are with Carnegie Mellon University, Pittsburgh, PA, 15213, Email: {dianam, radum}@cmu.edu*

optimization (MOO) problem.

In this work, we incorporate appropriate analytical models for each of the relevant objectives (*i.e.,* throughput, latency, temperature, and energy). We also demonstrate that it is necessary to consider all objectives to achieve the optimal trade-off between temperature and performance. We examine the differences between performance-only and performance-thermal-joint optimization as an example. However, the complexity of the design space and the high number of objectives make this design optimization problem difficult. Widely-used MOO techniques (*e.g.,* NSGA-II [9] or simulated annealing based AMOSA [10]) can require significant amounts of time due to their exploratory nature. Therefore, more efficient and scalable optimization techniques are required.

To this end, in this work, we propose a new MOO algorithm, MOO-STAGE, which extends the machine learning framework STAGE [11]. As opposed to traditional MOO algorithms that only consider the current solution set when making search decisions, MOO-STAGE learns from the knowledge of previous search trajectories to guide the search towards more promising parts of the design space. This significantly reduces the optimization time without sacrificing the solution quality. Using MOO-STAGE, we can take advantage of the traffic characteristics of different applications and incorporate appropriate design objectives to enable quick design space exploration of 3D heterogeneous systems. In addition, through careful analysis, we notice that several applications on heterogeneous platforms exhibit similar traffic patterns. Subsequently, we propose that an application-agnostic heterogeneous 3D NoC can be designed to achieve similar performance as designs that are optimized for a specific application. We evaluate the feasibility and performance of these application-agnostic designs across all considered benchmarks.

Below we summarize our main contributions in this work:

1.  We undertake a comprehensive study of the traffic patterns from multiple applications taken from various domains running on 3D heterogeneous systems.
2.  Based on the observed traffic patterns, we propose a generalized application-agnostic heterogeneous 3D NoC design that achieves similar levels of performance (latency, throughput, energy, and temperature) as application-specific designs.
3.  We propose a new MOO framework MOO-STAGE and apply it to the problem of manycore 3D heterogeneous NoC design. Our findings show that MOO-STAGE can find the same quality of solutions as AMOSA and a branch-and-bound based algorithm (PCBB [12]) while significantly reducing optimization time and improving scalability.

## 2 RELATED WORK

In this section, we present some of the most relevant prior works on 3D heterogeneous NoC design and related MOO algorithms.

### 2.1 3D Heterogeneous NoCs

Due to its heterogeneity, CPU-GPU based systems exhibit several interesting traffic characteristics, for instance, GPUs typically only communicate with a few shared last level caches (LLCs) which results in many-to-few traffic patterns (*i.e.,* many GPUs communicate with a few LLCs) with negligible inter-GPU communication [7], [13], [14]. This can cause the LLCs to become bandwidth bottlenecks under heavy network loads and lead to significant performance degradation [7]. In addition, since heterogeneous systems share the memory resources, the GPUs can monopolize the memory and cause high CPU memory access latency [15]. Conventional 2D architectures, such as mesh NoCs, cannot efficiently handle this many-to-few traffic or fulfill the quality of service (QoS) requirements for both CPU and GPU communication [7].

In recent years, designers have taken advantage of 3D IC's higher packing density and lower interconnect latency to improve the performance of manycore systems [4], [5]. The advantages of 3D integration for CPU and GPU based manycore systems have been demonstrated in [16], [17] where the authors have principally focused on improving the throughput and energy efficiency by using the benefits of 3D integration for homogeneous systems (all CPUs or all GPUs) only.

Due to the differences in the thread-level parallelism of CPUs and GPUs, the NoC designed for heterogeneous systems should satisfy *both* CPU and GPU communication constraints [18]. Hence, designing the 3D NoC for heterogeneous systems is more complicated than homogeneous systems; this aspect has not been explored adequately. On top of this, 3D ICs suffer from thermal issues due to higher power density [8], [19]. One of the common methodologies for reducing the peak temperature in a 3D architecture includes proper core placement to prevent high power consuming cores from being placed on top of each other [19]. However, it is not possible to implement such a strategy for heterogeneous systems with many GPU cores [8]. Other techniques to reduce temperature include suitable floorplanning [20] and temperature-aware task scheduling [21]. In contrast to these prior works, we propose a MOO algorithm to intelligently place the cores and links within a 3D heterogeneous system that jointly considers all relevant design metrics, *e.g.,* latency, throughput, energy, and temperature.

For a given workload, application-specific NoCs are known to outperform conventional architectures, *e.g.,* mesh NoCs [7]. A MOO formulation for 3D NoCs is presented in [8] for accelerating deep learning workloads. In [22], the authors have explored heterogeneous NoC design for multimedia applications. However, these works have only focused on one class of workloads to design the NoC and ignored the correlation in the traffic patterns of other applications.

### 2.2 Multi-Objective Optimization Algorithms

Basic MOO algorithms such as genetic algorithms (GA), *e.g.,* NSGA-II [9], or simulated annealing-based algorithms, *e.g.,* AMOSA [10], have been used in different optimization problems. AMOSA has been demonstrated to be superior to GAs or simulated annealing [10] and has been applied for the problem of heterogeneous NoC design in [7], [8]. However, since AMOSA is based on simulated annealing, it needs to be annealed slowly to ensure a good solution, which does not scale well with the size of the search space.

TABLE 1
LIST OF APPLICATIONS AND THEIR RESPECTIVE DOMAINS

| Applications | Domain/Usage |
|---|---|
| Back Propagation (BP) | Pattern Recognition |
| Breadth-First Search (BFS) | Graph Algorithm |
| CNN for CIFAR -10 (CDN) [28] | Image Classification (RGB) |
| Gaussian Elimination (GAU) | Linear Algebra |
| HotSpot (HS) | Physics Simulation |
| CNN for MNIST (LEN) [27] | Image Classification (Grayscale) |
| LU Decomposition (LUD) | Linear Algebra |
| Needleman-Wunsch (NW) | Bio-Informatics |
| k-Nearest Neighbors (KNN) | Data Mining |
| PathFinder (PF) | Grid Traversal |

In [23], the authors have used a heuristic-based MOO for multicore designs. However, they focus mainly on optimizing individual cores in smaller systems with up to 16 processors. Latency and area have been optimized using GAs to design NoC architectures in [24]. The authors in [25] have used machine learning techniques like linear regression and neural networks for MOO on different platforms. A learning-based fuzzy algorithm has been proposed to reduce the search time in [26]. However, this methodology requires a threshold to be decided for each application separately. A recent work [12] proposed a branch-and-bound-based algorithm, priority and compensation factor-oriented branch and bound (PCBB) for task mapping in a NoC-based platform [12]. However, this work only considers task mapping on a relatively smaller system size, where calculating the bound for each node is significantly easier. These works have mainly considered homogeneous platforms with smaller system sizes and fewer number of objectives.

3D heterogeneous NoC design is far more complex since the design must consider the requirements for each component. With additional constraints such as temperature and energy, the required optimization time can become tremendously high. Therefore, as systems become more complex, algorithms that are scalable with the size of the search space and can reduce optimization time without sacrificing solution quality will be needed.

In this work, we show that multiple applications exhibit similar traffic patterns on heterogeneous platforms. Leveraging this observation, we investigate the design of application-agnostic NoC architectures and propose a machine-learning inspired algorithm MOO-STAGE for 3D heterogeneous NoC design. Together, using MOO-STAGE and our observations of application traffic characteristics, we significantly reduce the design time of 3D heterogeneous NoCs and create optimized, application-agnostic architectures.

## 3 TRAFFIC PATTERN ANALYSIS

In this section, we present an in-depth study of the characteristics of the traffic patterns generated by a variety of applications that run on a heterogeneous platform. To this end, applications from multiple domains were selected, *e.g.*, physics, data mining, and bio-informatics. Two of these benchmarks, LeNet [27] and CDBNet [28], are commonly used neural networks for image classification while the rest of these applications come from the Rodinia benchmark suite [29]. This allows us to study the traffic patterns and the corresponding communication requirements of commonly used applications from different fields. Table 1 lists the applications along with their corresponding domains/usages. To obtain accurate traffic characteristics, we run each application on a detailed architecture simulator, Gem5-GPU [30]. The traffic characteristics are measured in the number of flits per cycle. Full experimental details are elaborated in Section 6.1.

Fig. 1 shows the traffic heat map for BP, BFS, NW, and PF applications running on a generic 64-tile heterogeneous system (8 CPUs, 16 LLCs, and 40 GPUs). Each row represents a different source, while each column represents a different destination. Since CPU and GPU cores have different requirements for delivering high performance, we show their respective traffic patterns separately, CPU-LLC communication in the top section and GPU-LLC communication in the bottom section (of note, CPU-GPU communication is negligible).

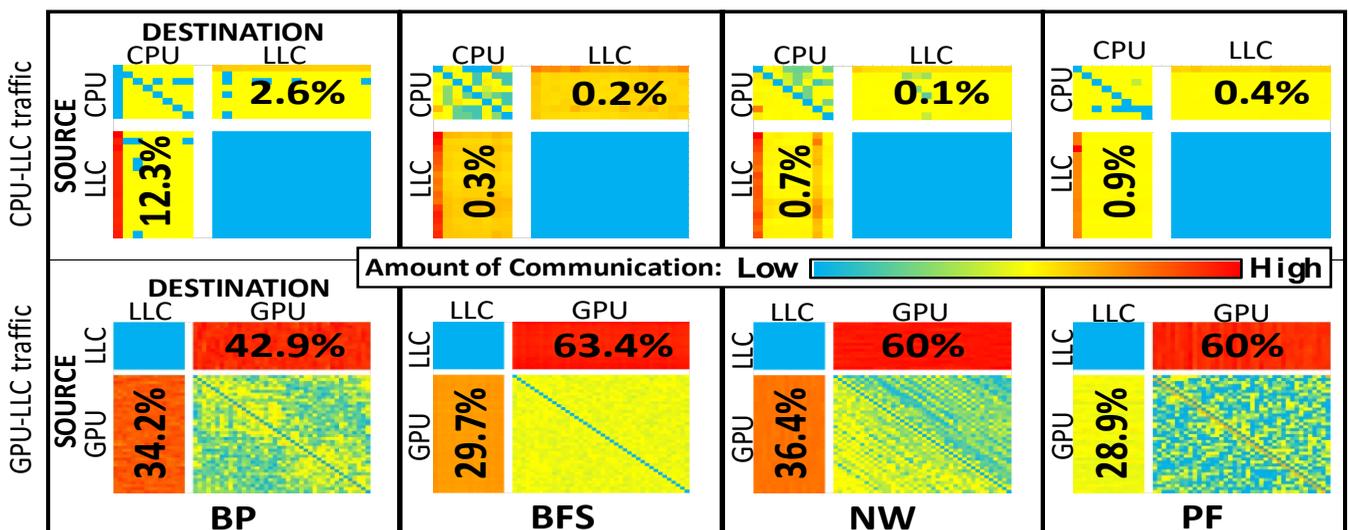

Fig. 1. Traffic pattern heat map for four applications (BP, BFS, NW, and PF) running on a 64-tile heterogeneous manycore system. The numbers indicate percentage of total traffic contributed by that section (*e.g.*, CPU-LLC communication results in 2.6% of total traffic for BP).

We observe that these heterogeneous systems exhibit several interesting traffic patterns:

- In every application, we observe that one CPU (the master core) exhibits higher traffic intensity compared to the other CPUs. The master core is easy to spot since it contributes a majority of the CPU traffic.
- Contrary to CPUs, GPU-LLC pairs exhibit nearly uniform high traffic due to well distributed and parallelized GPU workloads. The large number of GPUs can cause the GPU traffic to significantly congest the network. Communication between the other pairs of cores, *e.g.*, GPU-GPU is much lower.
- The majority of all traffic is associated with the LLCs. Fig. 2 shows percentage of total traffic going to/from the LLCs. On average, more than 80% of the total traffic is associated with the LLCs. Since heterogeneous systems typically have a small number of LLCs, this generates *many-to-few* communication patterns, especially between the GPUs and LLCs [7]. Without proper architectural support, LLCs can easily become network hotspots.

All applications considered (Table 1) exhibit such traffic behaviors and have similar traffic heat maps. Based on these observations, we conjecture that these characteristics are more dependent on the heterogeneous architecture than any specific application. Even though there exists some amount of application-specific variations among the cores, these differences are relatively insignificant compared to the heavy many-to-few communication going to/from the LLC blocks. As a result, the traffic patterns of any new application can be expected to exhibit similar features as those in Fig. 1. Therefore, an NoC optimized for any of these applications can potentially be re-used for other applications without significant loss of performance.

To demonstrate that the above-mentioned traffic patterns are not specific to any particular system size, we consider a different system size of 36 tiles (4 CPUs, 8 LLCs, and 24 GPUs). The traffic patterns generated by this system size exhibit the same characteristics as the 64-tile system across all applications, *i.e.*, a highly active master core, little CPU-GPU communication, nearly uniform GPU-LLC communication, and most of the communication is based around the LLCs (Fig. 2). We do not replicate Fig. 1 for the 36-tile system for brevity. This reinforces our previous observation that the traffic characteristics are more dependent on the elements of the heterogeneous architecture and is not limited to one system size or configuration. Hence, we can design the 3D NoC architecture by primarily considering the constituents of the heterogeneous system rather than any specific traffic patterns.

# 4 MULTI-OBJECTIVE OPTIMIZATION FORMULATION

## 4.1 Drawbacks of Mesh NoCs

Mesh NoC is the preferred design for on-chip communication due to its simplicity. Intel's Xeon Phi and Tilera's TILE processors are examples of architectures with a mesh NoC. However, as the number of cores on a single chip increases, mesh NoCs inevitably require more hops for each network traversal. These added hops lead to increased network latency and energy consumption. Therefore, despite its simplicity, mesh NoCs do not scale well with system size.

Mesh NoCs are especially ill-suited for heterogeneous systems. In [7], the authors have shown that links closer to the LLCs are highly over-utilized due to the many-to-few communication in mesh NoCs. Even an optimized 3D mesh can have links carrying 3X the average link traffic [8]. This can lead to network congestion, which results in higher latency and decreased throughput, negatively affecting the overall system performance. To combat these issues, we look to define a general methodology for designing 3D NoC-based heterogeneous systems.

## 4.2 MOO formulation for 3D heterogeneous NoCs

In this section, we discuss the necessary objectives to design an efficient 3D heterogeneous system. Fig. 3 illustrates an example 3D heterogeneous architecture with two layers. For these systems, it is important that we 1) optimize both CPU and GPU communication; 2) efficiently balance the load of the 3D NoC under many-to-few traffic patterns seen in Section 3; 3) minimize the network energy; and 4) minimize the peak temperature of the system. The design methodology should optimize the system for individual core requirements along with other design constraints for high-performance NoC architectures. There may be additional design objectives based on specific design cases which can be similarly included in the design process. In this work, the design methodology focuses on the placement of the CPUs, GPUs, LLCs, and planar links. We elaborate on how the methodology satisfies each objective next.

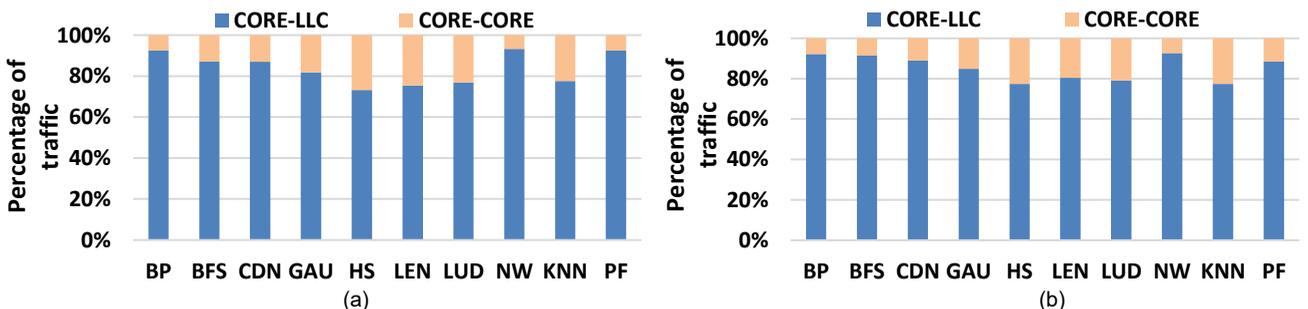

Fig. 2. Traffic breakdown showing the percentage of traffic between (in either direction) LLC and either CPU or GPU (CORE-LLC) and between CPUs and GPUs (CORE-CORE) for a (a) 36-tile and (b) 64-tile manycore system.

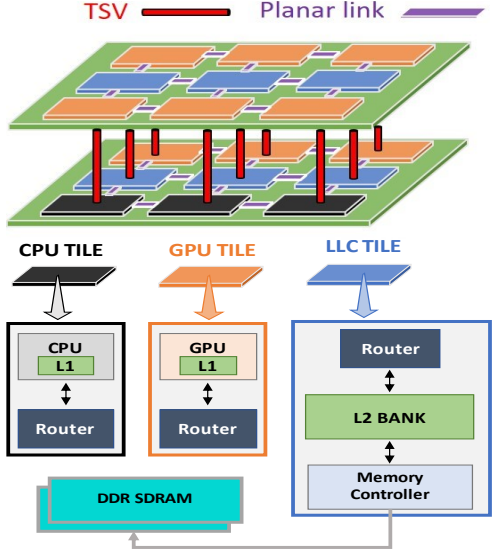

Fig. 3. Overview of the TSV-based 3D system considered in this work. The system is divided into CPU, GPU, and LLC tiles. Tiles are interconnected via a planar link (intra-layer) or a TSV (inter-layer). This figure is for illustration purpose only; it is not optimized for any metric.

It should also be noted that the objectives considered here are generic and can be used for homogeneous 3D NoC design as well. For example, in a GPU-only system, we should optimize throughput while for a CPU-only system we should optimize the overall latency. Apart from these, other design objectives considered in this work, e.g., energy, temperature are also applicable in a homogeneous setting.

### 4.2.1 CPU Communication Objective
CPU cores use instruction-level parallelism to achieve high performance on a limited number of threads. If any of these threads stall, CPUs incur a large penalty. Therefore, memory access latency is a primary concern for CPUs. For $C$ CPUs and $M$ LLCs, we model the average CPU-LLC latency using the following equation [5]:

$$Lat = \frac{1}{C*M} \sum_{i=1}^{C} \sum_{j=1}^{M} (r \cdot h_{ij} + d_{ij}) \cdot f_{ij} \qquad (1)$$

Here, $r$ is the number of router stages, $h_{ij}$ is the number of hops from CPU $i$ to LLC $j$, $d_{ij}$ indicates the total link delay, and $f_{ij}$ represents the amount of interaction between core $i$ and core $j$. The path from core $i$ to core $j$ is determined by the routing algorithm (given in Section 6.1). It should be noted here that the above equation is not limited to our specific routing technique and can be used with other routing algorithms as well.

### 4.2.2 GPU Communication Objective
Unlike the CPUs, GPUs rely on high levels of data parallelism. Massive amounts of parallelism coupled with quick context switching allow the GPU to hide most of its memory access latency. However, to do so, GPUs need lots of data and rely on high throughput memory accesses.

We maximize the throughput of GPU-related traffic by load-balancing the network to allow more messages to utilize the network at a time. In other words, given a frequency of traffic interaction and the routing paths, we want to balance the expected link utilization across all links. This does not change the total number of packets to be communicated. Instead, it reduces the number of heavily congested links by redistributing traffic flows. This reduces the amount of contention for heavily utilized links. As a result, links are more readily available, there is less network congestion, and hence, network throughput is improved.

For more intuition, load-balancing the network by adjusting link and tile placement tries to bring highly communicating tiles closer together and place links such that path diversity between highly communicating pairs is created. In other words, this load-balancing approach attempts to improve throughput by utilizing the given resources more efficiently. To balance the expected link utilization (load-balance the network), we consider minimizing both the mean ($\overline{U}$) and standard deviation ($\sigma$) of expected link utilization as suitable objectives.

The expected utilization of link $k$ ($U_k$) can be obtained by the following equation:

$$U_k = \sum_{i=1}^{R} \sum_{j=1}^{R} (f_{ij} \cdot p_{ijk}) \qquad (2)$$

Here $R$ is the total number of tiles and $p_{ijk}$ indicates whether a planar/vertical link $k$ is used to communicate between core $i$ and core $j$ respectively, *i.e.,*

$$p_{ijk} = \begin{cases} 1, & if\ cores\ i, j\ communicate\ along\ planar/vertical\ link\ k \\ 0, & otherwise \end{cases}$$

$p_{ijk}$ can be determined by using the network connectivity and routing protocols.

Then, the mean ($\overline{U}$) and standard deviation ($\sigma$) of link utilization can be determined from the following equations:

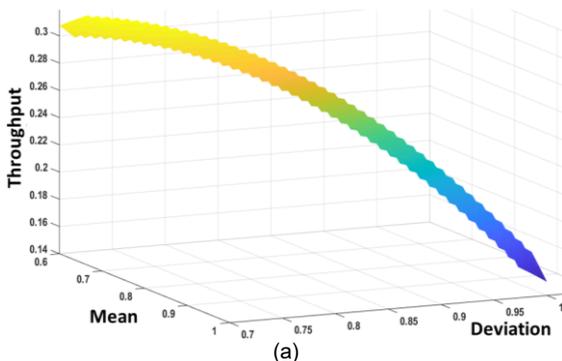

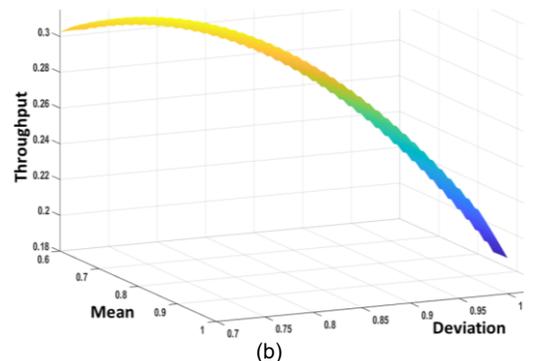

Fig. 4. Throughput with respect to mean (Eq. 3) and standard deviation (Eq. 4) of link utilization for (a) BFS and (b) HS. The plots have been generated by NoCs that were visited while optimizing for throughput *only* (Section 6.2, Case 1).

$$\bar{U} = \frac{1}{L}\sum_{k=1}^{L} u_k \tag{3}$$

$$\sigma = \sqrt{\frac{1}{L}\sum_{k=1}^{L}(u_k - \bar{U})^2} \tag{4}$$

**Model Validation:** Throughput can be accurately measured from network simulations. However, repeated simulations require significant amounts of time and increase total optimization time [26]. Existing network throughput models have only considered regular networks [31] and hence cannot be applied to the networks we generate (there are no regularity constraints). In this work, we have modeled maximizing throughput as minimizing Eqs. (3) and (4). We validate our proposed throughput model using detailed cycle-accurate network simulations. Fig. 4(a) and Fig. 4(b) show the throughput trend for different values of mean (Eqn. 3) and standard deviation (Eqn. 4) of link utilization for BFS and HS. Similar behavior is observed for all other applications. The plots have been restricted to regions which had enough data points for a faithful representation. It is clear from these figures that network throughput has an inverse relation with the mean and standard deviation of link utilization. Reducing the mean and standard deviation simultaneously leads to a monotonic increase in throughput. Therefore, increasing throughput can alternatively be expressed as minimizing mean and standard deviation of the expected link utilization, validating our throughput model.

### 4.2.3 Thermal requirements

One of the key challenges in 3D integration is the high-power density and resulting temperature hotspots. High temperature not only affects performance but also the lifetime of the device. Cores that are further away from the sink tend to have higher temperatures than those close to the sink. Therefore, cores must be properly placed, *e.g.*, high power consuming cores should be placed close to the sink to reduce temperature.

To estimate the temperature of a core, we use the fast approximation model presented in [32]. It considers both horizontal and vertical heat flow to accurately estimate the temperature. A manycore system can be divided into $N$ single-tile stacks, each with $K$ layers, where $N$ is the number of tiles on a single layer and $K$ is the total number of layers. The temperature of a core within a single-tile stack $n$ located at layer $k$ from the sink ($T_{n,k}$) due to the vertical heat flow is given by:

$$T_{n,k} = \sum_{i=1}^{k}\left(P_{n,i}\sum_{j=1}^{i}R_j\right) + R_b\sum_{i=1}^{k}P_{n,i} \tag{5}$$

This represents the vertical heat flow in a manycore system [32]. Here, $P_{n,i}$ is the power consumption of the core $i$ layers away from the sink in single-tile stack $n$, $R_j$ is the vertical thermal resistance, and $R_b$ is the thermal resistance of the base layer on which the dies are placed. The values of $R_j$ and $R_b$ are obtained using 3D-ICE [33]. The horizontal heat flow is represented through the maximum temperature difference in the same layer $k$ ($\Delta T(k)$):

$$\Delta T(k) = \max_n T_{n,k} - \min_n T_{n,k} \tag{6}$$

The overall thermal prediction model includes both vertical and horizontal heat flow equations. Following [32], we use $T$ as our comparative temperature metric for any given 3D architecture:

$$T = \left(\max_{n,k} T_{n,k}\right)\left(\max_k \Delta T(k)\right) \tag{7}$$

### 4.2.4 Energy requirements

A few long-range links added to the NoC can improve performance [5]. However, these long-range links are costlier in terms of energy. Routers with a higher number of ports can improve path diversity and throughput, however, larger routers are difficult to design and are power hungry. Therefore, router size and link length must be optimized during design time to deliver high performance without consuming high amounts of energy. For a system with $N$ tiles, $R$ routers, $L$ planar links, and $V$ vertical links, the approximate network energy consumed is obtained using the following equation.

$$E_{router} = \sum_{i=1}^{N}\sum_{j=1}^{N} f_{ij} \cdot \sum_{k=1}^{R} r_{ijk} \cdot (E_r \cdot P_k) \tag{8}$$

$$E_{link} = \sum_{i=1}^{N}\sum_{j=1}^{N} f_{ij} \cdot \left(\sum_{k=1}^{L} p_{ijk} \cdot d_k \cdot \mathrm{E}_{planar} + \sum_{k=1}^{V} q_{ijk} \cdot E_{vertical}\right) \tag{9}$$

$$E = E_{router} + E_{link} \tag{10}$$

Here $E_r$ denotes the average router logic energy per port and $P_k$ denotes the number of ports available at router $k$. The total link energy can be divided into two parts due to the different physical characteristics of planar and vertical links. $f_{ij}$ represents the frequency of communication between core $i$ and core $j$ that can be extracted from Gem5-GPU simulations while $d_k$ represents the physical link length of link $k$. Here, $q_{ijk}$ and $r_{ijk}$ is defined similarly as $p_{ijk}$ (Eqn. 2) to indicate if a vertical link or router $k$ is utilized to communicate between core $i$ and core $j$ respectively. $E_{planar}$ and $E_{vertical}$ denote the energy consumed per flit by planar metal wires and vertical links respectively. All the required power numbers were obtained using Synopsys Prime Power for 28nm nodes. The total network energy $E$ is the sum of router logic and link energy.

### 4.2.5 Overall MOO formulation

In the end, our aim is to find a 3D heterogeneous manycore design that minimizes the mean link utilization ($\bar{U}$), standard deviation of individual link utilizations ($\sigma$), average latency between CPU and LLCs ($Lat$), temperature ($T$) and energy ($E$). It is important to note that the analytical models for these objectives only need to be accurate in determining which designs are better *relative* to one another, *e.g.*, lower values of $T$ result in better temperatures. This allows us to quickly compare designs without performing detailed simulations during the optimization search. Since optimizing one objective may negatively affect another, it is important that these objectives are optimized simultaneously. For example, a thermal-only aware placement would move high-power cores closer to the sink [8] and possibly further away from cores they highly communicate with, negatively affecting performance and energy. We write our combined objective as follows:

$$D^* = MOO(OBJ = \{\bar{U}(d), \sigma(d), Lat(d), T(d), E(d)\}) \tag{11}$$

where, $D^*$ is the set of Pareto optimal designs among all possible 3D heterogeneous manycore system configurations $D$, *i.e.*, $D^* \in D$, $MOO$ is a multi-objective solver, and $OBJ$ is the set of all objectives to evaluate a candidate design $d \in D$. A candidate design $d$ consists of an adjacency matrix for the links (designates which pair of tiles are connected via a link) and a tile placement vector (designates which core is placed at which tile). We also ensure that for all $d \in D$, all source-destination pairs have at least one path between them. Since mesh is the most commonly used NoC architecture, any design $d$ has an equal number of links as that of a 3D mesh NoC.

In the following section, we describe the machine learning based MOO-STAGE, that we use as the multi-objective problem solver. However, any other MOO algorithm can also be used.

# 5 DESIGN OPTIMIZATION USING MACHINE LEARNING

In this section, we present a machine learning based optimization algorithm called MOO-STAGE that is scalable with the size of the search space. STAGE [11] is an online learning algorithm originally developed to improve the performance of local search algorithms (*e.g.*, hill climbing) for single objective optimization problems. In [5], authors have shown that STAGE can significantly outperform traditional optimization techniques, namely, simulated annealing (SA) and genetic algorithms (GA) for NoC design optimization with homogenous cores.

Inspired by this success of STAGE for single objective NoC design optimization, we extend it to a multi-objective optimization setting. In this work, we propose *MOO-STAGE*, a multi-objective optimization algorithm, and apply it to 3D manycore heterogeneous NoC design. The key idea behind MOO-STAGE is to intelligently explore the search space such that the MOO problem is efficiently solved. More precisely, MOO-STAGE utilizes a supervised learning algorithm that leverages past search experience (*Local search*) to learn an evaluation function that can then estimate the outcome of performing a local search from *any* given state in the design space (*Meta search*). In practice, the MOO-STAGE algorithm iteratively executes *Local* and

---

**Algorithm 1.** Local Search: $local(Obj, d_{start})$

**Input:** $Obj$ (Set of optimization objectives),
$\qquad d_{start}$ (Starting design)
**Output:** $S_{local}$ (Non-dominated set of designs),
$\qquad S_{traj}$ (Trajectory set), $d_{last}$ (Last design)

1:   **Initialize:** $S_{local} \leftarrow \{d_{start}\}$, $S_{traj} \leftarrow \{d_{start}\}$,
$\qquad d_{curr} \leftarrow d_{start}$
2:   **While** 1:
3:      $d_{next} \leftarrow \arg\max\limits_{d \in neigh(d_{curr})} PHV_{Obj}(S_{local} \cup \{d\})$
4:      **If** $PHV_{Obj}(S_{local} \cup \{d_{next}\}) > PHV_{Obj}(S_{local})$:
5:         $\big|\; S_{local} \leftarrow S_{local} \cup \{d_{next}\}$
$\qquad\qquad S_{local} \leftarrow \{d \in S_{local} | (\nexists k \in S_{local})[k \prec d]\}$
6:      **Else:**
7:         $\big|\;$ **Return** $(S_{local}, S_{traj}, d_{last} \leftarrow d_{curr})$
8:      $d_{curr} \leftarrow d_{next}$
9:      $S_{traj} \leftarrow S_{traj} \cup \{d_{curr}\}$

---

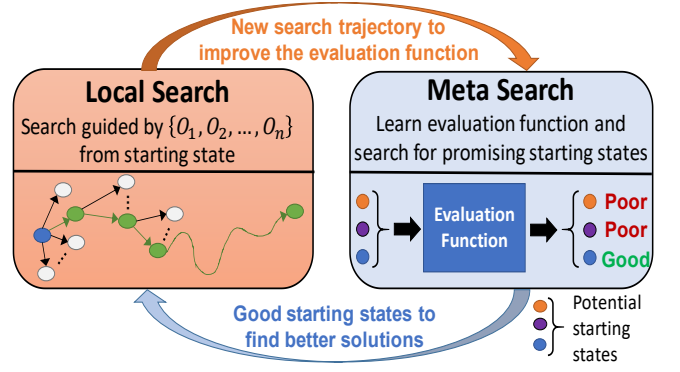

Fig. 5. Overview of the MOO-STAGE algorithm.

*Meta* searches in a sequence as shown in Fig. 5.

Fig. 5 shows a high-level overview of how MOO-STAGE works. The first stage (*Local search*) performs a search from a given starting state, guided by a cost function considering all objectives. Then, the search trajectories collected from the *Local search* is used for the next stage (*Meta search*) to learn an evaluation function. This evaluation function attempts to learn the potential (quantified using the cost function) of performing a *Local search* starting from a particular state. This allows the algorithm to prune away bad starting states to reduce the number of *local search* calls needed to find (near-) optimal designs in the given design space. Unlike MOO-STAGE, other MOO algorithms based on random restarts do not leverage any such knowledge and spend significant time searching from states that would otherwise be rejected by MOO-STAGE. Therefore, MOO-STAGE *explicitly* guides the search towards promising areas of the search space much faster than conventional MOO algorithms. Below we describe the details of the MOO-STAGE algorithm.

## 5.1 MOO-STAGE: Local Search

Given an objective, the goal of a local search algorithm (*e.g.*, greedy search or SA) is to traverse through a sequence of neighboring states to find a solution that minimizes the objective. To accommodate multiple objectives, we employ the Pareto hypervolume (PHV) [34] metric to evaluate the quality of a set of solutions (higher is better). The PHV is the hypervolume of the dominated portion of the objective space as a measure for the quality of Pareto set approximations [34]. A design $P$ is dominated by design $Q$ ($P \prec Q$) when

$$\forall i : Obj_i(P) \leq Obj_i(Q) \wedge \exists i : Obj_i(P) < Obj_i(Q)$$

Local search guided by the PHV heuristic has two strong advantages over other metrics for comparing solutions [35]: 1) The PHV captures the improvement in any objective. If a new set of solutions has a better PHV than the current set of solutions, then the new set of solutions covers more of the objective space and better captures the trade-offs between objectives. 2) PHV allows the handling of *any* number of objectives as part of the MOO problem (*i.e.*, generality) since PHV maps to a single output (cost). This is particularly useful for learning the evaluation function via a regression learning algorithm.

To compute the PHV, we employ a fast and scalable PHV algorithm called hypervolume by slicing objectives

| **Algorithm 2.** MOO-STAGE |
|---|
| **Input:** $Obj$ (Set of optimization objectives), $iter_{max}$ (Maximum iterations), $D$ (Design space) |
| **Output:** $S_{global}$ (Non-dominated set of designs) |

| | |
|---|---|
| 1: | **Initialize:** $S_{global} \leftarrow \emptyset, S_{train} \leftarrow \emptyset, d_{start} \leftarrow rand(D)$ |
| 2: | **For** $i = 0$ to $iter_{max}$: |
| 3: | $\left( S_{local}, S_{traj}, d_{last} \right) \leftarrow local(Obj, d_{start})$ |
| 4: | Maintain non-dominated global set: $S_{global} \leftarrow S_{global} \cup S_{local}$ $S_{global} \leftarrow \{ d \in S_{global} \big| (\nexists k \in S_{global})[k \prec d] \}$ |
| 5: | **If** $S_{global} \cap S_{local} = \emptyset$: [If algorithm converged] |
| 6: | **Return** $S_{global}$ |
| 7: | Add training example for each design $d \in S_{traj}$: $S_{train} \leftarrow S_{train} \cup \left\{ \left( d, PHV_{Obj}(S_{traj}) \right) \right\}$ |
| 8: | Train evaluation function: $Eval \leftarrow train(S_{train})$ |
| 9: | Greedy Search: $d_{restart} \leftarrow greedy(Eval, d_{last})$ |
| 10: | **If** $d_{last} = d_{restart}$: |
| 11: | $d_{start} \leftarrow rand(D)$ |
| 12: | **Else** |
| 13: | $d_{start} \leftarrow d_{restart}$ |
| 14: | **Return** $S_{global}$ |

[36]. It employs the divide-and-conquer principle to achieve efficiency: it repeatedly divides the PHV computation into simpler problems with fewer objectives and aggregates the solutions of simpler problems to compute the total hypervolume.

In this work, we use a simple greedy search with the objective of maximizing PHV with respect to the input objective set ($PHV_{Obj}$) as the local search procedure (**Algorithm 1**). However, it should be noted that greedy search has been employed as an example case only. Any other local search method, *e.g.*, SA, can be used to similar effect. Starting from the initial state $d_{start}$, we find the best neighboring state ($neigh(d_{curr})$) that improves the PHV heuristic at each greedy search step (**Algorithm 1**, line 3). In the context of designing 3D heterogeneous systems, a neighboring state is where exactly one planar link is repositioned or two tiles are swapped (both irrespective of layers). If this best neighboring state improves the PHV value, we add this state to the set of local optima ($S_{local}$) while ensuring that all designs in $S_{local}$ are non-dominated (**Algorithm 1**, lines 4-5). This is repeated until the best neighboring state does not improve the PHV value, at which point, we return the local optima set, search trajectory ($S_{traj} = d_{start}, ..., d_{last}$), and the final search state ($d_{last}$). *Essentially, the local search explores the neighborhood of the current solutions to expand the Pareto front to dominate as much of the objective space as possible.*

### 5.2 MOO-STAGE: Meta Search

The second and key component of MOO-STAGE is the learning phase, also known as the meta-search. For standard local search procedures, one of the key limitations is that the quality of the local search critically depends on the starting point of the search process ($d_{start}$). Although algorithms like SA try to mitigate this effect by incorporating some random exploration, they are still limited by the *local* nature of the search. If the search repeatedly begins near poor local minima, it is possible that the search will never find a high-quality solution. MOO-STAGE attempts to solve this problem by learning a function approximator (evaluation function) using previous *local search* data that can predict the outcome of a *local search* procedure from a particular starting point. Using this evaluation function MOO-STAGE intelligently selects starting states with a high potential to lead to better quality solutions and subsequently, significantly reduces the computation time. We discuss the details of this procedure in the following paragraphs.

After completing the local search, we add the local optima set to the global optima set ($S_{global}$) ensuring that all states in the global optima set are non-dominated (**Algorithm 2**, lines 3-4). If the local optima set didn't add any new entries to the global optima set, MOO-STAGE completes and returns the global optima set (**Algorithm 2**, lines 5-6). Otherwise, we add the local search trajectory ($S_{traj}$) and PHV of this trajectory ($PHV_{Obj}(S_{traj})$) as a training example to the aggregated training set ($S_{train}$) and learn the evaluation function $Eval$ using $S_{train}$ (**Algorithm 2**, lines 7-8). In this work, we employ Regression Forest as the base learner for creating $Eval$. Regression Forest is only used as an example here and other regression learners that are quick to evaluate and sufficiently expressive to fit the training data can be used to similar effect.

Given the function $Eval$, we use a standard greedy search to optimize $Eval$ beginning at the last state of the local search ($d_{last}$) to find the starting state for the next local search iteration ($d_{restart}$). If $d_{last}=d_{restart}$, we choose a random design from the design space instead ($rand(D)$) (**Algorithm 2**, lines 9-13). Using these two computational search processes (*Local search* and *Meta search*), MOO-STAGE progressively learns the structure of the solution space and improves $Eval$. *Essentially, the algorithm attempts to learn a regressor that can predict the PHV of the local optima from any starting design and explicitly guides the search towards predicted high-quality starting designs.*

## 6 EXPERIMENTAL RESULTS

### 6.1 Experimental Setup

To obtain network- and processor-level information, we use the Gem5-GPU full-system simulator [30]. The CPU cores are based on the x86 architecture while the GPUs are based on the NVIDIA Maxwell architecture. Here, each GPU core is analogous to a Streaming Multiprocessor (SM) in Nvidia terminology. Within each GPU core, we have 32 shader processors. The architecture of an individual GPU core is similar to a GPU Compute Unit (CU) described in [30]. The CPUs operate at 2.5 GHz while the GPUs operate at 0.7 GHz. The core power profiles have been extracted using GPUWattch [37] and McPat [38]. The core temperatures have been obtained using 3D-ICE [33]. Due to the high-power densities in 3D ICs, we incorporate microfluid-based cooling techniques to reduce core temperatures. In this work, we also adopt Reciprocal Design Symmetry (RDS) floor-planning [20] to reduce the direct overlap of core areas as much as possible.

To implement different NoC topologies, we modified the Garnet network [39] in Gem5-GPU. It should be noted that Garnet [39] includes a detailed cycle-accurate interconnection network model that incorporates appropriate flow control, effects of link/buffer contention, etc. In this work, we use a standard three-stage router, however, the proposed design methodology is independent of the number of router stages. The 3D mesh NoCs use XYZ-dimension order routing while the proposed architectures use ALASH routing [40]. It should be noted here that the proposed architectures do not have a regularity constraint and hence XYZ-dimension order routing cannot be employed as in the case of 3D Mesh NoCs. The memory system uses a MESI Two-Level cache coherence protocol. Each CPU and GPU have a private L1 data and instruction cache of 32 KB each. Each LLC consists of 256KB memory.

To evaluate our proposed MOO-STAGE, we consider two reference algorithms, AMOSA [10] and PCBB [12]. AMOSA is a widely employed algorithm for multi-objective optimization due to its ability to achieve near optimal solutions [10]. On the other hand, PCBB is a recently proposed branch-and-bound based technique used for task mapping in an NoC-based system considering multiple objectives [12]. PCBB outperforms standard branch and bound techniques due to two key features: a) a prioritization strategy that prioritizes more prominent tasks to help prune branches earlier in the process and reduce computational complexity; and b) a compensation factor that allows tradeoffs between bound computational overhead and accuracy. We have adapted PCBB for heterogeneous 3D NoC design as follows. First, we divide the branching decisions into two stages, node placement followed by link placement. Second, we estimate the bound of a branch using a roll-out procedure by virtually placing the remaining unplaced cores and links following several well-known mapping strategies (*i.e.*, greedy, random, and small-world). Lastly, similar to [12], we combine the objectives into a single metric. We prune a branch only if the bounds are worse even after being adjusted by a compensation factor, indicating that the branch is unlikely to produce a good solution even after accounting for bound estimation error [12].

We evaluate the algorithms based on runtime and quality of solutions. Given the set of Pareto-optimal solutions $D^*$ specified by (11) for each MOO solver considered here (*i.e.*, AMOSA, PCBB, and MOO-STAGE), we run detailed simulations on this subset of solutions to get absolute values for energy, performance, and temperature. Here, the NoC solution is characterized using network energy-delay product (EDP) as an example. The network EDP is a combined metric for performance and energy. Here, network

EDP is defined as the product of network latency and energy consumption. All experiments have been run on an Intel® Xeon® CPU E5-2620 @ 2GHz machine with 16 GB RAM running CentOS 6. The code for MOO-STAGE, AMOSA, and PCBB have been made available on GitHub [41].

## 6.2 Optimization Parameters

System designers often have many different and perhaps conflicting objectives. Therefore, we look at several cases with different number of objectives for the proposed 3D heterogeneous architecture. As an example, we consider three different cases:

**Case 1** – $\{\bar{U}, \sigma\}$ We consider mean (Eq. 3) and standard deviation of link utilization (Eq. 4).

**Case 2** – $\{\bar{U}, \sigma, Lat\}$ We add CPU-LLC latency (Eq. 1) to Case 1.

**Case 3** – $\{\bar{U}, \sigma, Lat, E\}$ We add energy (Eq. 10) to Case 2. However, new objectives can be judiciously added to fit the designer goals and constraints.

Also, since both AMOSA and MOO-STAGE rely on the structure of the design space, we define what constitutes a neighboring design. In the context of designing 3D heterogeneous systems, a neighboring state is where exactly one planar link is repositioned or two tiles are swapped (can be between tiles in the same or different layers). On the other hand, since PCBB is based on branch-and-bound, it does a systematic enumeration of the candidate solutions. Finally, our goal here is to optimize the placement of CPUs, LLCs, GPUs, and planar links such that they improve the design objectives.

## 6.3 Finding Better Solutions Using Machine Learning

In this section, we investigate PCBB, AMOSA, and MOO-STAGE's performance for the problem of 3D heterogeneous NoC design. More specifically, we investigate their abilities to optimize the core and planar link placement for 3D heterogeneous architectures. Here, we consider a 64-tile system with 8 CPUs, 16 LLCs, and 40 GPUs. The number of planar links and number of TSVs are kept the same as a similar size 3D mesh NoC. When comparing the two algorithms, we present the average EDP of multiple runs from the same starting NoC configuration.

For brevity, in Fig. 6 we present the BFS results averaged over multiple runs as a representative example. Similar observations are made for all other applications as well. Fig. 6 shows the evolution of the best solution's EDP over time for AMOSA and MOO-STAGE for all three optimization cases. Since PCBB is not an anytime algorithm, we can only show the total run-time needed to complete the branch-and-bound enumeration. This is discussed later in Table 2.

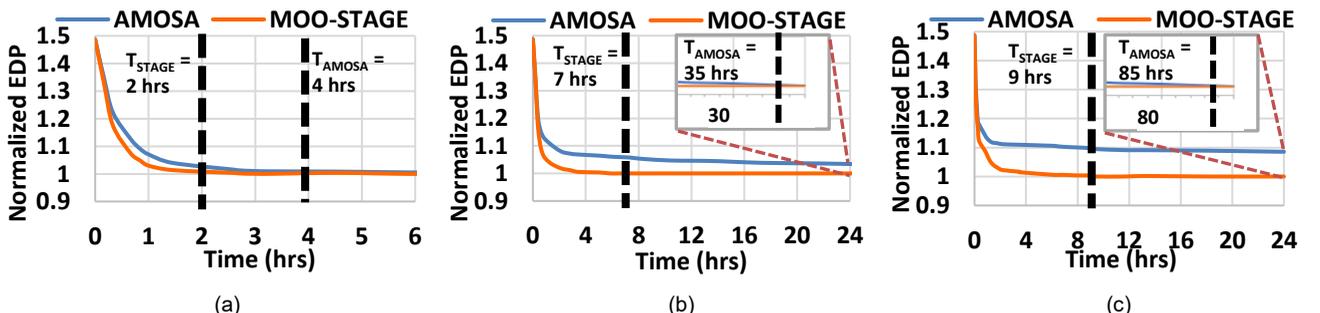

Fig. 6. Normalized quality of NoC solutions (EDP) obtained using AMOSA and MOO-STAGE for (a) two objectives ($\{\bar{U}, \sigma\}$), (b) three objectives ($\{\bar{U}, \sigma, Lat\}$), and (c) four objectives ($\{\bar{U}, \sigma, Lat, E\}$) for the BFS benchmark.

TABLE 2
MOO-STAGE SPEED-UP OVER PCBB AND AMOSA

| Application | Two-obj | | Three-obj | Four-obj |
|---|---|---|---|---|
| | PCBB | AMOSA | AMOSA | AMOSA |
| BP | 130 | 1.5 | 6.4 | 12.5 |
| BFS | 135 | 2.0 | 5.0 | 9.4 |
| CDN | 146 | 1.5 | 5.8 | 13.7 |
| GAU | 134 | 1.3 | 6.0 | 7.2 |
| HS | 144 | 1.5 | 8.0 | 10.0 |
| LEN | 145 | 2.0 | 5.8 | 14.2 |
| LUD | 140 | 1.3 | 5.0 | 10.0 |
| NW | 150 | 1.5 | 5.0 | 11.4 |
| KNN | 148 | 1.2 | 6.4 | 7.5 |
| PF | 142 | 1.2 | 5.0 | 11.4 |
| **Average** | **141.4** | **1.5** | **5.8** | **10.7** |

It is evident from Fig. 6 that MOO-STAGE achieves lower EDP values significantly faster than AMOSA. To further demonstrate this, we define two metrics: $T_{MOO-STAGE}$ which is the time required for MOO-STAGE to converge and $T_{AMOSA}$ which is the time needed for AMOSA to generate similar quality of solutions. However, AMOSA never finds the best solution that MOO-STAGE obtains even after significantly longer durations for the three- and four-objective optimization. For these cases, $T_{AMOSA}$ is defined as the time AMOSA takes to reach within 3% of the best solution quality of MOO-STAGE in terms of EDP. It is clear from Fig. 6 that the amount of speed-up MOO-STAGE achieves increases as the number of objectives increase. With four objectives, MOO-STAGE converges approximately after $T_{MOO-STAGE} = 9$ hours while AMOSA takes approximately $T_{AMOSA} = 85$ hours to come within 3% of MOO-STAGE's solution quality. Therefore, MOO-STAGE achieves an approximate 9.4 times optimization time speed-up compared to AMOSA.

The significant improvement in optimization time can be attributed to the fact that MOO-STAGE performs *active learning*. In machine learning literature, it is well-known that the active learning paradigm is exponentially more efficient than passive supervised learning [42]. Similar to other active learning algorithms, *e.g.*, DAgger [43], MOO-

STAGE aggregates learning examples over multiple iterations to reduce the number of training data needed to learn a target concept. This guarantees that only a small number of trajectories are needed to achieve good generalization behavior with the learned function [43] and accurately evaluate the entire input design space. As a result, after a few iterations MOO-STAGE achieves a near-accurate evaluation function to speed-up the optimization process.

Table 2 shows the speed-up with MOO-STAGE compared to AMOSA and PCBB for all applications under Cases 1, 2, and 3 (Section 6.2). MOO-STAGE achieves significant gains in convergence time for all applications and number of objectives. Note that due to the large execution time for PCBB, we only show the two-objective optimization case (Case 1) for PCBB. However, increasing the number of objectives will reduce the number of branches that are pruned and exponentially increase the run-time. This would result in even worse three- and four-objective run-times for PCBB.

As seen from Table 2, even for the simpler two-objective optimization, PCBB takes 141x longer on average to find the similar quality of solution as MOO-STAGE. This is mainly due to the sheer size of the design space of 3D NoCs. For more intuition, in a 4x4x4 (64-tile) system with 144 links (96 planar + 48 vertical), the total number of possible tile placements is 64 factorial. Then, each of these tile placements has $C(C(16,2) * 4,96)$ different ways to place the planar links. Although PCBB manages to prune significantly over 99.99% of this solution space, the tiny fraction that is left consists of several millions of possible solutions. This is significantly more than MOO-STAGE or AMOSA leading to worse execution times.

On the other hand, MOO-STAGE reduces the optimization time over AMOSA by 1.5X, 5.8X, and 10.7X on average for the two-, three-, and four-objective cases respectively. Table 2 also shows that, MOO-STAGE can obtain high-quality solutions in a shorter amount of time irrespective of the application. For further analysis, Fig. 7 shows the tile placement and number of planar links in each layer for the NoC configurations obtained using MOO-STAGE and AMOSA at time $T_{STAGE}$. The BFS benchmark considering four-objectives (Case-3 described in Section 6.2) is shown as an example. Here, we do not consider using PCBB as it takes orders of magnitude more time to generate a good NoC design compared to both MOO-STAGE and AMOSA. From Fig. 7, we note that in the NoC obtained using MOO-STAGE, LLCs tend to remain in the middle layers. This allows the LLCs to access the vertical links in both directions and reduce the average hop count to the other tiles. Also, we observe that more links are concentrated in layers with LLCs. The presence of more links enables greater path diversity and reduces the amount of traffic congestion under many-to-few traffic pattern leading to better performance. Of note, it is interesting to see that AMOSA and MOO-STAGE both achieve nearly similar tile placement configurations but very different link placements. This is due to the link placement search space being much larger than the tile placement search space. Therefore, AMOSA fails to explore it adequately within $T_{STAGE}$ time and ends up with a solution similar to the initial starting NoC, i.e., 3D Mesh (Starting state for all searches is 3D Mesh which has uniform distribution of links among layers as well). These analyses justify the approximately 9.5% difference in EDP (Fig. 6(c)) between the best solutions obtained using the two algorithms at time $T_{STAGE}$. Therefore, by learning the search space, MOO-STAGE is able to achieve better quality solutions in a shorter span of time.

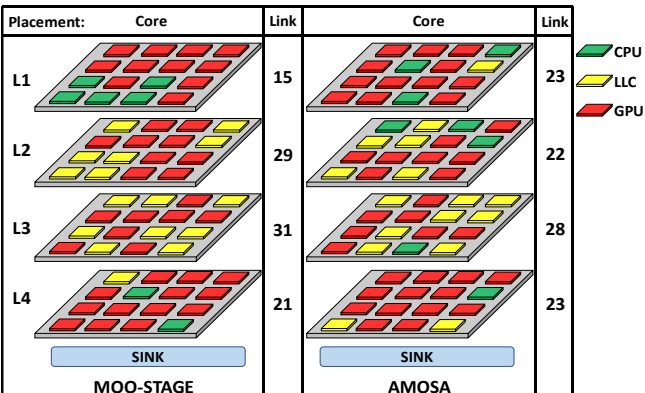

Fig. 7. Comparison of physical placement of tiles and planar links in solutions obtained using MOO-STAGE and AMOSA.

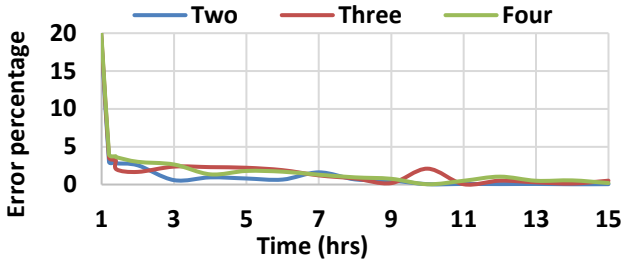

Fig. 8. Prediction error for MOO-STAGE (64-tile system, BFS).

To demonstrate MOO-STAGE's ability to learn a function that accurately maps the design space to the objective space, we show the prediction error of the evaluation function *Eval* (**Algorithm 2**, lines 7-8) as a function of time for all considered cases (Section 6.2, Cases 1-3) in Fig. 8 considering BFS as an example. The prediction error (in %) represents the difference between the estimated PHV value obtained by *Eval* and the actual PHV value obtained by the subsequent local search. From Fig. 8, we note that irrespective of the number of objectives, after only a few hours, the prediction error is less than 5%. This low error rate indicates that the evaluation function *Eval* can accurately predict good starting points for the local search. Hence, MOO-STAGE continuously improves its search by choosing promising starting points. As seen in Fig. 6 and Table 2, this allows MOO-STAGE to reduce the total number of searches necessary and find solutions much more quickly than AMOSA's *relatively* random explorations and PCBB's systematic enumeration of the entire candidate set.

In Section 3, we studied the traffic patterns generated by different applications. We found that the traffic patterns of different applications on heterogeneous platforms exhibit a set of similar characteristics. Therefore, we conjectured that we could utilize a heterogeneous platform optimized for one application to run other applications. Taking advantage of the similarities in application traffic characteristics seen in Section 3, we undertake the design of application-agnostic NoC architectures using MOO-STAGE in the next sections.

## 6.4 Application-Agnostic NoC Design

In this section, we validate our observations and show that NoCs optimized for one application can show similar performance for other applications as well. Here, we consider the four-objective optimization problem (Section 6.2, Case 3) as an example to reduce network energy and CPU-LLC latency, while improving the GPU-LLC throughput. So, principally we focus on enhancing the network efficiency (performance) here. To show our approach's applicability to different system sizes, we consider the optimization of a 36-tile (4 CPUs, 8 LLCs, and 24 GPUs arranged in four layers of 3x3 cores) and a 64-tile system (8 CPUs, 16 LLCs, and 40 GPUs arranged in four layers of 4x4 cores).

To design an application-agnostic NoC, we consider two cases: generate an NoC optimized for a) each application (denoted by its application) and b) a set of several applications, using an aggregated traffic profile (AVG). For each of the $N$ applications, we create a different AVG NoC (a total of $N$ AVG NoCs) using the set of remaining $N − 1$ applications (leave-one-out).

The optimized NoCs are then used to execute all applications, *e.g.*, an NoC optimized for *BFS* is used to execute all ten applications, and the performance is normalized to the application's respective application-specific NoC. For example, the EDP of an NoC optimized for *BFS* running *BP* is normalized with respect to the EDP of the NoC optimized for *BP*, running *BP*. Each AVG NoC executes the application that was left-out during optimization (otherwise unknown to the optimization). Fig. 9(a) shows normalized EDP of 64-tile NoCs. From Fig. 9(a) we note that on average, only 3.2% degradation is observed for all applications when compared to application-specific NoCs with a worst case reaching only up to 9.8%. However, the averaged NoC (AVG) only shows a 1.1% average degradation compared to the application specific NoC architectures.

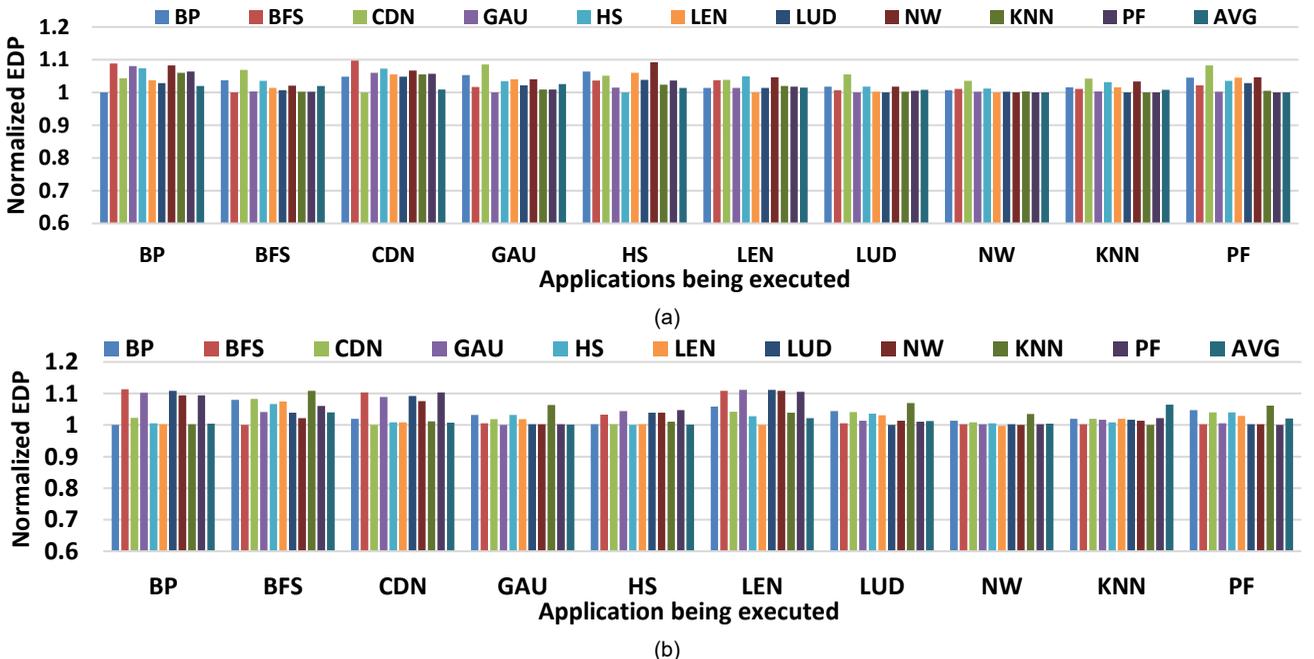

(a)

(b)

Fig. 9: Normalized EDP of (a) 64-tile and (b) 36-tile NoCs optimized for *network efficiency only (*Section 6.2, Case 3*)* to study the performance degradation with respect to application specific designs.

In Fig. 9(b) we also provide a comparative study with a 36-tile system. Here, we see similar evaluation results for the 36-tile system as well. From Fig. 9(b), we note that even for a different system size, the performance degradation is only 3.8% on average, with worst case difference going up to 11% for NoCs optimized with a single application. Similar to previous case, AVG performs better with an average degradation of 1.8%.

By aggregating the characteristics from multiple applications, AVG can better generalize to the unknown application. Therefore, an NoC optimized for a subset of applications can be reused for a new application on 3D heterogeneous systems without significant performance penalty. These implications can be helpful for future applications and NoC designs. For example, an NoC optimized using *BFS* and *GAU* can be used to execute neural network architectures like *LeNet* or *CDBNet* as shown in Fig. 9(a). Similarly, other neural network architectures, *e.g.*, *AlexNet* [44], is likely to exhibit similar performance improvements. Hence, irrespective of system size, it is possible to design high-performance 3D heterogeneous NoCs without prior knowledge of the application we intend to run.

## 6.5 Thermal Aware Application-Agnostic NoC Design

Up until this point, we have only optimized 3D heterogeneous systems for network efficiency (network performance). However, 3D ICs have higher packaging densities, resulting in higher temperature. High on-chip temperature is detrimental to the performance of the IC. Hence, it is essential to include the thermal characteristics in the optimization process. In this section, we extend our evaluation of application-agnostic 3D heterogeneous NoC design by including temperature into the optimization process as well. We introduce two new optimization cases (extending from the cases in Section 6.2) for this purpose:

**Case 4 – {$T$}** Thermal only optimization. We consider peak core temperature (Eq. 7) only.

**Case 5 -** {$\overline{U}, \sigma, Lat, T, E$} Joint performance-thermal optimization. We add temperature (Eq. 7) to Case 3.

Like the previous section, we consider two system sizes with single application optimized NoC and the averaged NoC. However, optimizing only for the thermal profile can lead to performance degradation since it doesn't consider any performance objectives during the design process. We show the performance-thermal trade-offs in Fig. 10.

In Fig. 10, we compare the results of NoCs optimized for Case 3 (network efficiency/performance), Case 4 (thermal-only), and Case 5 (joint network performance-thermal) normalized to the Case 3 NoC. Figs. 10(a) and 10(b) show the Full-system execution time and EDP respectively. Fig. 10(c) shows the temperature of the 3D NoC configuration in all three NoC cases. Here, Full-System EDP (FS-EDP) is defined as the product of Full-System execution time and Energy. The full-system execution time is obtained via detailed Gem5-GPU simulations. It is clear from Fig. 10 that incorporating only thermal in the optimization process leads to the best temperature profile but a significant degradation of more than 7% in full-system execution time on average. Similarly, the NoC optimized for network efficiency (Case 3) achieves the best EDP but at a 20ºC average degradation of temperature compared to the only-thermal optimized NoC (Case 4). On the other hand, the jointly-optimized NoC exhibits temperature improvements of 18ºC on average while sacrificing only 2.3% in overall execution time. Therefore, it is important that we jointly optimize both performance and thermal to reduce on-chip temperature while delivering high performance.

Next, we show that it is also possible to design application-agnostic NoC architectures for jointly-optimized thermal-performance case. To this end, we perform similar experiments as in Fig. 9. Fig. 11(a) (64-tile system) and Fig 11(b) (36-tile system) show the normalized EDP for applications executed on different application-specific and traffic-averaged NoCs. Exactly like Fig. 9, the application-specific NoC for each application has been chosen as the baseline for comparison. On average, only 2.8% degradation is observed for the application-specific NoC running other applications, when compared to the application-specific NoCs on its application, with a worst case of 8.5%. Similarly, for the 36-tile NoCs, the average EDP degradation becomes 4.5% and worst case is 11%. Like the previous case, the traffic-averaged NoCs perform better with an average

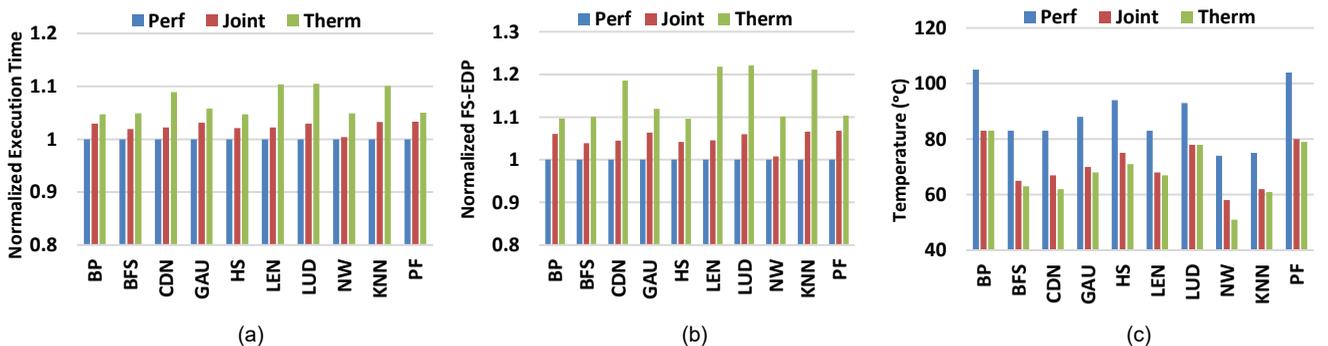

Fig. 10: Performance-thermal trade-offs for 64-tile NoCs: (a) Full-System Execution time, (b) Full-System EDP, (c) Temperature comparison for three optimization cases: network efficiency/performance-only (Perf), joint performance-thermal (Joint) and thermal-only (Therm).

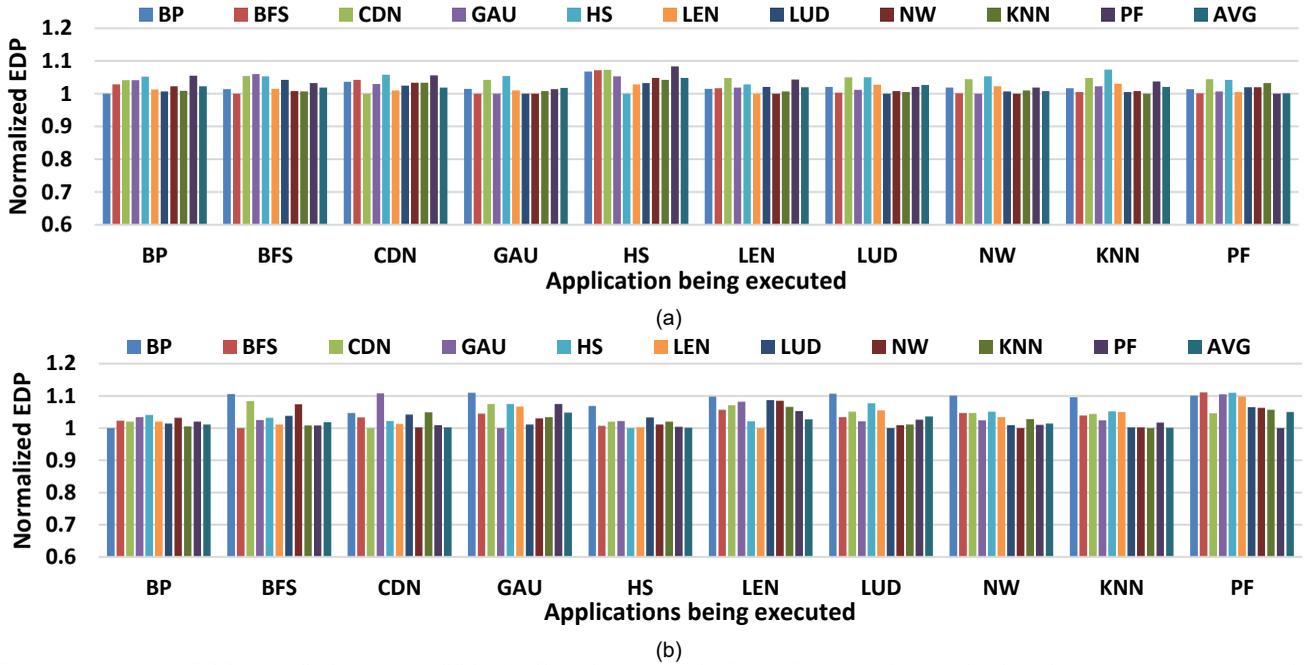

Fig. 11: Normalized EDP of (a) 64-tile and (b) 36-tile NoCs optimized jointly for *performance-thermal* to study the performance degradation with respect to application-specific designs.

degradation of 2% and 2.1% for 64-tile and 36-tile NoCs respectively.

From the above observations, we find that due to the similarities in the traffic pattern of applications on a heterogeneous platform, it is possible to optimize the NoCs for any known application(s) and have them perform well with unknown applications. We have seen that optimizing on a small set of applications reduces both the average and worst-case degradation even further. Looking deeper, we study the physical core and link distributions for each of the application-specific NoCs. In Section 3, we noted that the traffic patterns are similar for multiple applications. As a result, the optimized NoCs are expected to be similar as well.

To this end, we evaluate the link distribution among the four layers and the associated tile placements for the heterogeneous NoCs. Fig. 12 shows the distribution of tiles and links in the performance-only optimized *Het-perf* (Section 6.4), joint performance-thermal optimized *Het-joint* (Section 6.5), and *Mesh-perf* (3D Mesh NoC with tile placement that has been performance-only optimized similar to *Het-perf*). Please note that *Het-perf* in Fig. 12 is the same

NoC obtained using MOO-STAGE shown in Fig. 7 and is repeated here to easily observe the differences among NoCs optimized considering different objectives. Due to the uniform link distribution across all layers, mesh NoCs cannot handle many-to-few traffic efficiently (Section 4.1). On the other hand, both heterogeneous NoCs designed following the framework presented in Section 4 produce an irregular topology with more links near the LLCs. Compared to *Het-perf*, the placement of cores and links are greatly affected by doing a temperature-aware optimization (Fig. 12). Unlike *Het-perf* (which places LLCs mostly in the middle two layers), to reduce core temperatures, high power consuming cores, *i.e.*, GPUs, are placed closer to the sink in *Het-therm*. As a result, the LLCs and CPUs are placed mostly in the upper layers. Also, similar to *Het-perf*, more links are observed in the layers with a higher number of LLCs. In both these cases, the physical distribution of cores and links are observed to be similar for all considered applications. Hence, the optimized NoCs share similar characteristics in the physical placement of cores and links as well.

# 7 CONCLUSIONS

3D NoC-enabled CPU-GPU based heterogeneous architectures provide an opportunity to design high-performance, energy-efficient computing platforms to meet the growing computational need in deep learning and big-data applications. However, 3D heterogeneous architectures present several new design challenges: a) multiple potentially conflicting design requirements; b) 3D integration induced thermal hotspots; and c) significantly larger design spaces.

In this work, we have shown that we can generate thermally-efficient high-performance 3D NoCs that are application-agnostic by analyzing the on-chip traffic, designing suitable objectives, and using efficient MOO techniques.

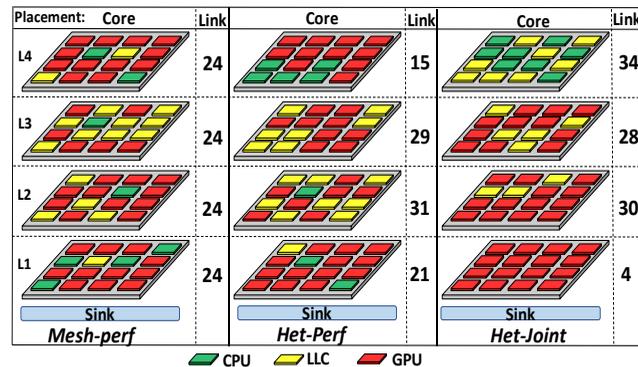

Fig 12: Distribution of tiles and links in different architectures considered in this work.

Our study shows that applications on heterogeneous systems with many GPUs and few LLCs, generate similar traffic patterns. Experiments demonstrate that our design framework can generate generic 3D NoC configurations which experience an average performance loss of 1.1% for 64-tile systems and 1.8% for 36-tile systems compared to application-specific NoCs by considering an aggregated traffic pattern of several applications. Similar observations were made for a performance-thermal joint optimized case. These observations were made irrespective of system size, system configuration, and available training application sets, demonstrating that we can create NoCs that generalize well to unknown applications using a small subset of available applications.

## ACKNOWLEDGMENT

This work was supported in part by the US National Science Foundation (NSF) grants CNS-1564014, CNS-1564022, CCF 1514269 and USA Army Research Office grant W911NF-17-1-0485.

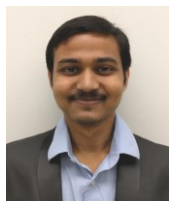

**Biresh Kumar Joardar** is a PhD student at Washington State University, Pullman, under the guidance of Dr. Partha Pratim Pande and Dr. Janardhan Rao Doppa. His current research interests include novel interconnect architectures for multicore chips, near memory computing and machine learning for electronic design automation.

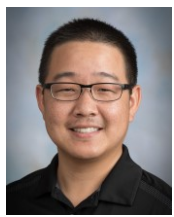

**Ryan Gary Kim** is an assistant professor in the Electrical and Computer Engineering Department at Colorado State University, Fort Collins. His current research interests include electronic design automation techniques for scalable, fully-adaptive manycore systems and domain-specific architectures. He is a member of the IEEE.

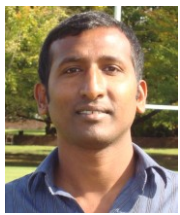

**Janardhan Rao Doppa** is an assistant professor in the School of Electrical Engineering and Computer Science at Washington State University. His research interests include machine learning and data-driven science and engineering with a special focus on electronic design automation and computer architecture. He received his PhD in computer science (2014) from Oregon State University.

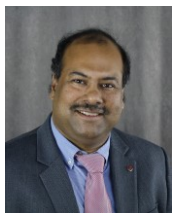

**Partha Pratim Pande (M'05, SM'11)** is a Professor and holder of the Boeing Centennial Chair in computer engineering at the school of Electrical Engineering and Computer Science, Washington State University, Pullman, USA. His current research interests are novel interconnect architectures for multicore chips, on-chip wireless communication networks.

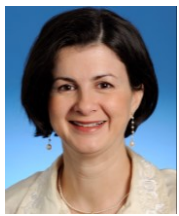

**Diana Marculescu** is a Professor of Electrical and Computer Engineering at Carnegie Mellon University. She has won several best paper awards in top conferences and journals in the area of low power design and design automation. Her research interests include sustainable and energy-aware computing, and computing for sustainability and other applications. She is an IEEE fellow.

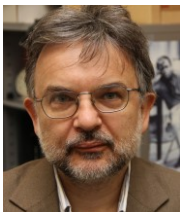

**Radu Marculescu** is a professor in the ECE Department at Carnegie Mellon University. He has received several Best Paper Awards in top conferences and journals covering design automation of integrated systems and embedded systems. His current research focuses on modeling and optimization of embedded and cyber-physical systems. He is a fellow of the IEEE.